\documentclass[draft]{agujournal2019}
\usepackage{url} 
\usepackage[a4paper,
            bindingoffset=0.2in,
            left=1in,
            right=1in,
            top=1in,
            bottom=1in,
            footskip=.25in]{geometry}
\usepackage{ragged2e}
\justifying
\usepackage{soul}
\usepackage{lineno}
\usepackage{hyperref}
\usepackage{fancyhdr}
\usepackage{amsmath}
\usepackage{wrapfig}
\usepackage{mathtools}
\usepackage{amsfonts}
\usepackage{amssymb}
\usepackage{textcomp}
\usepackage{booktabs}                      
\usepackage{multicol}
\usepackage{multirow}
\usepackage{xspace}
\usepackage{nomencl}
\usepackage{caption,booktabs}
\usepackage{float}
\usepackage{rotating}
\makenomenclature
\usepackage{tabularx}                      
\usepackage{natbib}
\usepackage{adjustbox}
\usepackage[graphicx]{realboxes}
\usepackage[toc,page]{appendix}
\usepackage{cancel}
\usepackage{graphicx}
\usepackage{epstopdf}
\usepackage{color}
\draftfalse

\newcommand{\revA}[1]{{\color{black}#1}}

\newcommand{\revB}[1]{{\color{black}#1}}

\journalname{}

\newcommand{\nsc}[1]{{\color{black}#1}}

\begin{document}

%
%


\title{Growth rate and energy dissipation \\ in wind-forced breaking waves}

%
%



%
\authors{
  Nicolò Scapin\affil{1,2},
  Jiarong Wu\affil{1,3},
  J. Thomas Farrar\affil{4},
  Bertrand Chapron\affil{5},
  Stéphane Popinet\affil{6}, and
  Luc Deike\affil{1,2}  
}
\affiliation{1}{Department of Mechanical and Aerospace Engineering, Princeton University, Princeton, NJ 08544, USA}
\affiliation{2}{High Meadows Environmental Institute, Princeton University, Princeton, NJ 08544, USA}
\affiliation{3}{Courant Institute of Mathematical Sciences, New York University, US}
\affiliation{4}{Woods Hole Oceanographic Institution (WHOI)}
\affiliation{5}{IFREMER, Univ. Brest, CNRS, IRD, Laboratoire d'Océanographie Physique et Spatiale (LOPS), France}
\affiliation{6}{Institut Jean Le Rond d’Alembert, CNRS UMR 7190, Sorbonne Université, Paris 75005, France.}
%
%
%
\correspondingauthor{Luc Deike}{ldeike@princeton.edu}
%
%
\begin{keypoints} 
  \item We perform high-fidelity simulations of wind-forced breaking waves and analyze the wind energy input, separating growth and breaking
  \item Wave growth scales as $(u_\ast/c)^2$ under wind forcing, with \revB{strong modulation due to wave slope} following non-separated sheltering
  \item Wave breaking triggers turbulence with a $z^{-1}$ dissipation profile, with magnitude controlled by the strength of the breaking event
\end{keypoints}
%
%
%

%
%


\begin{abstract} 
We investigate the energy growth and dissipation of wind-forced breaking waves at high wind speed using direct numerical simulations of the coupled air–water Navier–Stokes equations. A turbulent wind boundary layer drives the growth of a pre-existing narrowband wave field until it breaks, transferring energy into the water column. Under sustained wind forcing, the wave field resumes growth. We separately analyze energy transfers during wave growth and breaking-induced dissipation. \revB{Energy transfers are dominated by pressure input during growth and turbulent dissipation during breaking.} Wind input during growth is balanced with dissipation during breaking over an entire growing-breaking cycle. The wave growth rate scales with $(u_\ast/c)^2$, modulated by the wave steepness due to sheltering, and the energy dissipation follows the inertial scaling with wave slope at breaking, confirming the universality of the process. Following breaking, near-surface vertical turbulence dissipation profiles scale as $z^{-1}$, with their magnitude controlled by the breaking-induced dissipation.
\end{abstract}

\section*{Plain Language Summary} 
Ocean waves grow and eventually break under the action of strong winds, a process that plays a key role in transferring energy from the atmosphere to the ocean and limiting wave heights. In this study, we used high-resolution computer simulations to examine how wind causes waves to grow and how breaking waves transfer energy into ocean currents under high wind speed conditions. Waves grow mainly due to wind pressure on their surface, but once they become steep enough, they break and generate underwater turbulence that carries energy into the ocean. We confirm that the rate at which waves grow and break depends on their steepness and the wind strength. By analyzing the energy lost during breaking, we confirm a universal dissipation pattern that is independent of the specific mechanism leading to breaking. Finally, we propose a new way to describe how this energy is transferred below the surface, which agrees well with both our simulations and open ocean field measurements.
%
%
%
%
\section{Introduction}\label{sec:intro}
Ocean waves play a critical role in mediating energy transfer between the ocean and the atmosphere. The wind injects energy into the wave field, promoting wave growth and steepening. Once a critical amplitude or velocity is reached, waves break, dissipating energy and transferring it into the water column, generating currents and turbulence ~\citep{melville1985momentum,lamarre1991air,melville1996role,veron2001experiments,sutherland2015field}. 
Understanding wave growth rates and breaking-induced dissipation under strong wind forcing is critical to understand momentum and heat fluxes relevant to tropical cyclone intensification~\citep{sroka2021review}; while the associated energy fluxes control turbulence generation, which controls gas exchange~\citep {deike2022mass} and upper ocean currents~\citep{sullivan2010dynamics}. \par 
Regarding wave growth, two main theories describe different stages of the generation mechanism: the early stage, where growth is driven by turbulent pressure fluctuations~\citep{phillips1957generation,perrard2019turbulent,li2022principal}, and the finite-amplitude stage, where sufficiently steep waves modulate the atmospheric boundary layer~\citep{jeffreys1925formation,miles1957generation,belcher1993turbulent}. In the latter regime, the wind input scales with wave energy, theoretically leading to exponential wave growth. Field observations and experimental campaigns have reported the growth rate, expressed as a function of the ratio of the wind friction velocity and the wave phase speed $u_\ast/c$ with significant remaining scatter in the data~\citep{wu1979experimental,plant1982relationship,kihara2007relationship,grare2013growth,wu2022revisiting}. The role of wave steepness modulation at finite amplitudes and the accuracy of the sheltering framework from~\cite{belcher1993turbulent} using measured wind profiles during wave growth remains to be fully assessed. \par 
The total energy dissipation by wave breaking for individual breaking events  \citep{kendall1994energy,drazen2008inertial,banner2007wave} has been shown to scale with the wave amplitude at breaking, by extensive laboratory experiments without wind forcing and some experiments of breaking under wind forcing~\citep{grare2013growth}. Separately, field observations have discussed the vertical profile of turbulence dissipation rate and attempted to relate the integrated turbulence beneath an ensemble of breaking waves to the breaking statistics and estimations of the wind input into the wave field~\citep{gemmrich1994energy,craig1994modeling,terray1996estimates,zippel2022parsing,sutherland2015field,zippel2020measurements}. Local measurements have reported dissipation rates exceeding classical shear-layer predictions, with the excess attributed to the contribution of breaking events~\citep{terray1996estimates,sutherland2015field,zippel2022parsing,thomson2016wave,zippel2020measurements}. \par 
In this work, we unify the description of energy transfers, i.e. wind input and energy dissipation, between the wind and growing waves up to breaking. To quantify these processes under strong wind forcing, we perform fully resolved simulations of wind–wave interactions, including breaking, which builds upon our previous studies of momentum fluxes~\citep{scapin2025mom}, and distinguishes between the growing and breaking stages. We solve the general Navier-Stokes equations for two-phase air-water at high resolution, capturing scales ranging from millimeters to meters, without relying on subgrid models for waves or turbulence. This approach offers direct physical insights into wave breaking, informing subgrid-scale parameterizations for large-scale atmospheric and oceanic models. \par
This paper is organized as follows. Section~\ref{sec:dns_wfb} discusses the evolution of water currents, underwater velocity, and dissipation, and how these are influenced by wave breaking. Section~\ref{sec:en_flux} extracts the wave growth rate and quantitatively compares it with~\cite{belcher1993turbulent}, highlighting the role of wave steepness. Section~\ref{sec:diss} analyzes the breaking-induced dissipation, confirming the applicability of the inertial scaling from~\cite{drazen2008inertial} for wind-forced breaking waves. We show that vertical turbulence dissipation profiles transition to $\langle\varepsilon\rangle(z)\propto z^{-1}$ due to breaking. 
%
%
%
%
%
\section{Direct numerical simulation of wind-forced breaking waves}\label{sec:dns_wfb}
In this section, we briefly summarize the computational setup in \S\ref{sec:frame}, and in \S\ref{sec:evol}, we describe the evolution of the fully coupled system of wind, wave field, and underwater currents, illustrating the velocity and turbulence dissipation rate fields.
\begin{figure*}[h!]
  \centering
  \includegraphics[trim={0cm 0.0cm 0cm 0.5cm},width=\textwidth]{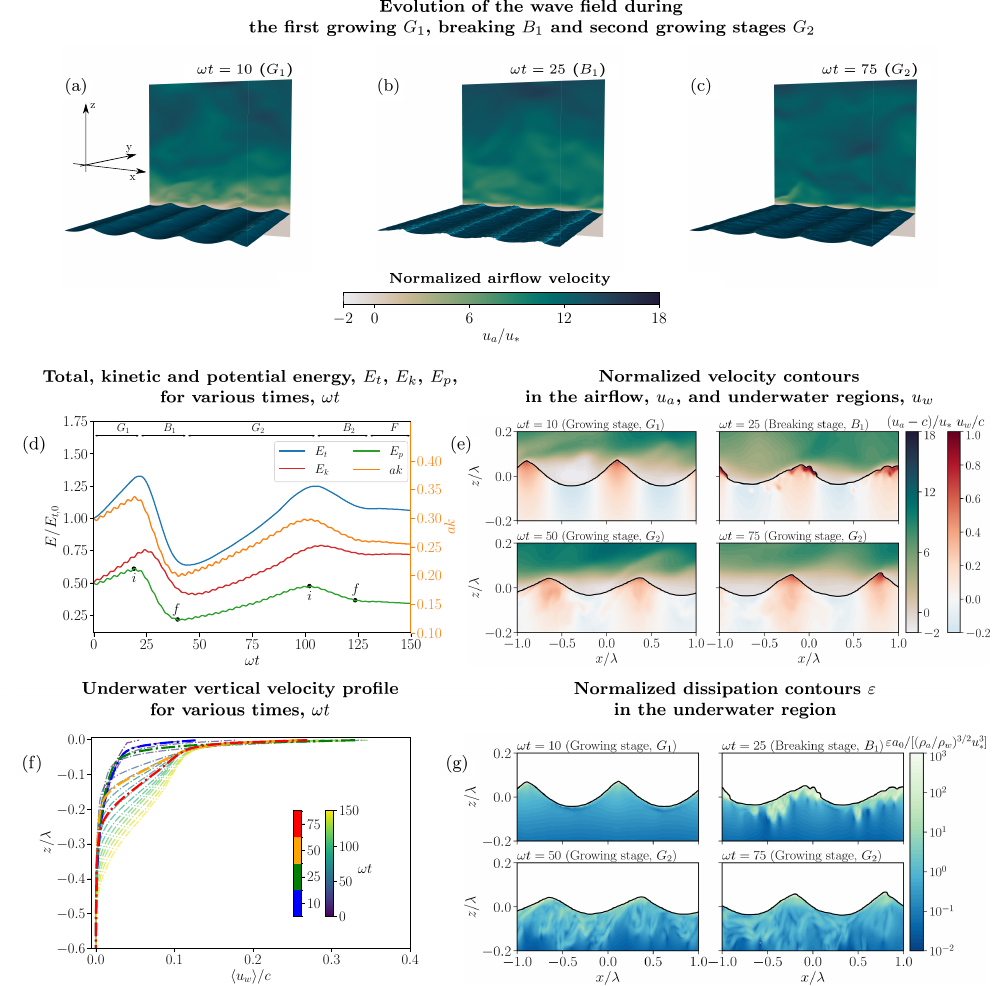}
  \caption{
  Evolution of the airflow, waves, and underwater currents for $u_\ast/c=0.9$. Top panels show snapshots of (a) first growing ($\omega t=10$, $G_1$), (b) breaking ($\omega t=25$, $B_1$), and (c) second growing ($\omega t=75$, $G_2$) stages under turbulent wind forcing along $x$, with contours of instantaneous streamwise velocity $u$ normalized by $u_\ast$.
  (d) Time evolution of potential ($E_p$), kinetic ($E_k$), and total ($E_t=E_p+E_k$) energy, all normalized by the initial total energy $E_{t,0}$. Equipartition of $E_t$ is lost due to the breaking event. The instantaneous steepness $a(t)k$ is shown on the right axis (orange curve).
  (e) Instantaneous streamwise velocity on the $y=0$ plane, normalized by $u_\ast$ (air) and $c$ (water), across the growing and breaking.
  (f) \revB{Vertical profiles of the mean streamwise velocity $\langle u_w\rangle$ in water, normalized by $c$. Profiles at $\omega t = [10,25,50,75]$ are shown in opaque colors, while profiles for $\omega t\in [0,150]$ with reduced transparency.}
  (g) Dissipation rate $\varepsilon(x,y,z)$ on the $y=0$ plane, normalized by $(\rho_a/\rho_w)^{3/2}u_\ast^3/a_0$.
  }
  \label{fig:fig1}
\end{figure*}

\subsection{Computational framework}\label{sec:frame}
We solve the two-phase air-water Navier-Stokes equations with surface tension using the open-source solver Basilisk~\citep{popinet2015quadtree,van2018towards,popinet2018numerical}. The numerical setup is provided in~\cite{wu2022revisiting,scapin2025mom}, and in the supplementary material.
The simulations fully resolve all relevant spatial and temporal scales, without relying on sub-grid models or prescribed wave motion. The computational domain is a cube of size $L_0$, spanning $[-L_0/2,L_0/2]^2 \times [-h_w,h_a]$, and is divided into two regions. The upper region $\Omega_a$ of height $h_a$ represents the turbulent wind (a fully developed turbulent boundary layer), while the lower region $\Omega_w$ of depth $h_w$ represents water currents, with densities $(\rho_a,\rho_w)$ and kinematic viscosities $(\nu_a,\nu_w)$. The domain is periodic in $x$ and $y$, with stress-free boundary conditions at $z=-h_w$ and $z=h_a$. The wave field is initialized at the interface $\Gamma$ using a third-order Stokes wave~\citep{deike2016air,mostert2022high}, with the corresponding irrotational underwater velocity field prescribed via potential theory~\citep{lamb1993hydrodynamics,deike2015capillary}. The fundamental wavelength is $\lambda$. Following~\cite{wu2022revisiting},~\cite{scapin2025mom}, the simulation parameters are $L_0=4\lambda$, $h_a=3.36\lambda$, and $h_w=L_0-h_a=0.64\lambda$, ensuring that the influence of the finite domain size on the wind boundary layer is negligible and that the wave propagation satisfies the deep-water dispersion relation. The air-water density ratio is $\rho_w/\rho_a = 816$. Five remaining non-dimensional groups govern the problem: the initial wave steepness $a_0k$ ($a_0$ the initial amplitude and $k=2\pi/\lambda$ the wavenumber), the ratio between the friction velocity $u_\ast$ and the wave speed $c$, i.e. $u_\ast/c$, the air and water Reynolds numbers, $Re_{\ast,\lambda}=u_\ast\lambda/\nu_a$ and $Re_W=c\lambda/\nu_w$, which reflects the importance of inertial over viscous forces in the air and in the water. Finally, the Bond number $Bo=(\rho_w-\rho_a)g/(\sigma k^2)$, where $g$ is gravity and $\sigma$ is the surface tension, represents the importance of gravity over surface tension forces. \par
Here, we fix $Re_W=2.55 \cdot 10^4$, $Bo=200$, and $a_0k=0.3$ to ensure that the wave field is in the gravity-dominated regime and that its growth is wind-driven~\citep{scapin2025mom}. For this study, we vary $u_\ast/c$ between $0.3$ and $0.9$, relevant to high-wind conditions like winter storms and tropical cyclones~\citep{sroka2021review}. $Re_{\ast,\lambda}$ is set to $214$ to maintain an inertial turbulent regime while keeping the simulation fully resolved. We also perform sensitivity analyses at $u_\ast/c=0.9$, exploring $Re_{\ast,\lambda}$ values from $54.5$ to $107$, and using a larger domain ($L_0=8\lambda$, $h_a=6.72\lambda$). These variations have negligible effects on energy fluxes.

\revA{Note that we use $Bo=200$ and we will obtain breaking waves with breaking slopes around $a k\approx 0.3$, the regime corresponds to spilling breakers~\citep{deike2015capillary}, for which bubble and droplet production remains moderate. Obtaining stronger breakers under wind forcing with substantial air entrainment and droplet emissions would require working with a more broadband wave field and a higher Bond number, leading to higher requirements in terms of numerical resolution to properly resolve the capillary and turbulence processes, see~\citep{deike2016air,mostert2022high}. However, as demonstrated by~\citep{deike2016air,mostert2022high,hao2024quantifying}, the dissipation due to breaking is nearly independent of the amount of bubbles and the $Bo$ and is dominated by the wave slope, as well as independent of $Re_W$ once $Re_W \gtrsim 20000$~\citep{deike2016air,mostert2022high,hao2024quantifying,di2022coherent}. The chosen parameters strike a balance to achieve fully resolved wind-wave breaking with energy dissipation representative of a high $Bo$ and high $Re$ regime at a manageable computational cost. Adaptive mesh refinement is used as described in \cite{scapin2025mom}, and the results presented here are verified to be independent of the grid size, as shown in the supplementary material.
}

\subsection{Development of the underwater currents}\label{sec:evol}
\revB{We analyze the development of underwater currents following wave-breaking events, starting from laminar and irrotational wave initial conditions, so that the transition to turbulence can be directly attributed to wave breaking.} \par 
The turbulent wind, propagating in the streamwise direction $x$ and visualized in the first three planes of figure~\ref{fig:fig1}, initially promotes the steepening and growth of the wave field, as shown in figure~\ref{fig:fig1}(a). When the critical threshold for breaking is reached, i.e. $(ak)_c \approx [0.28-0.33]$~\citep{banner1998tangential,perlin2013breaking,deike2015capillary}, the wave field breaks, as shown in figure~\ref{fig:fig1}(b). Once this stage is complete, the wave field grows again under wind forcing (fig.~\ref{fig:fig1}(c)). \par
These cycles of growing and breaking stages can be quantified by examining figure~\ref{fig:fig1}(d), which reports the wave potential energy $E_p$, the water kinetic energy $E_k$ and the total energy $E_t=E_p+E_k$ as a function of the dimensionless time $\omega t$ ($\omega=2\pi/T$ the angular frequency and $T=\lambda/c$ is the wave period). The kinetic energy is $E_k=\rho_w\int_{\Omega_w}|\mathbf{u}|^2/2 dV$, with $\mathbf{u}=(u,v,w)$ the velocity field, and the wave potential energy $E_p=\rho_wg\left(\int_{\Omega_w} z dV-(h_wL_0)^2/2\right)$ (a surface tension contribution to $E_t$ exists but is negligible owing to the large Bond number). \revA{Note that the potential energy $E_p$ is solely associated with the wave field, whereas the kinetic energy $E_k$ includes contributions from mean currents, turbulent fluctuations, and wave motion, and thus grows more rapidly than $E_p$ during the growing cycles.} Both $E_p$ and $E_k$ are shown in figure~\ref{fig:fig1}(d), normalized by the initial total energy, $E_{t,0}$, along with an estimate of the wave steepness $a(t)k$, plotted on the secondary y-axis as an orange curve. The wave steepness is estimated either using the amplitude defined as $a(t)=\sqrt{\int_\Gamma (2/\Gamma)(\eta - \overline{\eta})^2 d\Gamma}$, where $\eta(x,y)$ is the surface elevation and $\overline{\eta}$ its spatial mean, or from the local slope of the surface, i.e. $\max\left(\sqrt{(\partial\eta/ \partial x)^2 + (\partial \eta/\partial y)^2}\right)$. Both approaches were evaluated and yield very similar results~\citep{scapin2025mom}. \par
During the growing phases $G_1$-$G_2$, $E_p$ increases with wind input, while in breaking phases $B_1$-$B_2$ it decreases as energy transfers to the water column. In the final stage $F$, wind input and dissipation nearly balance~\citep{scapin2025mom}. Notably, $E_p$ during $G_2$ is lower than in $G_1$ due to reduced steepness $a(t)k$ (orange curve in figure~\ref{fig:fig1}(d)), while $E_k$ is higher from developing underwater currents and turbulence, which may also influence the wave growth~\citep{kudryavtsev2016growth}. \par
Throughout the wave field evolution, an underwater velocity current develops. Initially, during $G_1$, this current is mainly driven by wind stress (especially its viscous component) while the flow remains laminar, as shown in the first panel of figure~\ref{fig:fig1}(e). Around $\omega t \approx 20$, wave breaking occurs, transferring energy into the water column and perturbing the flow from the crest, where velocities approach the wave phase speed (second panel). During breaking, the flow accelerates and transitions to turbulence in the second growth cycle, as shown in the last two panels of figure~\ref{fig:fig1}(e). \par
This transition is also evident in the mean underwater velocity profile $\langle u_w \rangle$, computed as the spatial average over the $x$-$y$ plane using a wave-following coordinate~\citep{wu2022revisiting,scapin2025mom}, to capture the dynamics at both the crests and troughs of the wave field (see figure~\ref{fig:fig1}(f)). The profile becomes notably steeper near the wave field ($z/\lambda < 0.1$). Unlike $G_1$, currents during $G_2$ are driven by both wind stress and wave breaking, which triggers the transition to turbulence. Note that wave-following coordinates preferentially sample regions of higher positive velocity, an effect analogous to the origin of Stokes drift~\citep{pollard1973interpretation}. As a result, the velocity profile represents an Eulerian-Lagrangian combination. This averaging does not impact turbulence dissipation~\citep{wu2025turbulence}. \par
The transition to turbulence in the underwater currents is reflected in the turbulent dissipation, which is computed from the strain rate tensor as 
\begin{equation}\label{eqn:diss_loc}
  \varepsilon(x,y,z)=\nu_w(\partial_i u_j+\partial_j u_i)^2\mathrm{,}
\end{equation}
Figure~\ref{fig:fig1}(g) shows dissipation contours normalized by $(\rho_a/\rho_w)^{3/2}u_\ast^3/a_0$. During the initial growth stage $G_1$, $\varepsilon$ is nearly negligible and mainly due to viscous effects. \revB{When the wave breaks during $B_1$, dissipation increases and remains localized near the crest. After the breaking event, the wave field resumes its growth in $G_2$, a region of enhanced dissipation (relative to the previous growth cycle, $G_1$) develops deeper below the surface, reflecting the transition of water currents to turbulence induced by wave breaking. The transition to turbulence is here directly attributed to the breaking event given its very fast development over a time scale comparable to the wave period, while transition to turbulence resulting from the action of the wind stress arises on the order of hundreds of wave periods \citep{veron2001experiments}.}
%
%
%
%
\section{Energy fluxes and growth rate in wind-forced breaking waves}\label{sec:en_flux}
In this section, we study the energy exchanges between the airflow and the wave field: the energy flux due to wind input and the energy dissipation associated with wave breaking. The analysis is done by splitting the growing and breaking stages. In section~\ref{sec:wave_tmp}, we present the time evolution of the energy flux during the wave growing and breaking stages. In section~\ref{sec:growth_rate}, the non-dimensional wave growth rate due to the wind energy flux is discussed and compared to existing wind growth scaling theories, and we demonstrate the importance of the wave slope on growth through sheltering. 
\subsection{Temporal evolution of the wave energy budget}\label{sec:wave_tmp}
The energy transfer from turbulent airflow to the wave field is governed by the evolution of wave energy, defined as twice the potential energy, i.e. $E_W(t) = 2E_p(t)$, assuming equipartition. As discussed in section~\ref{sec:dns_wfb}, equipartition of total energy $E_t$ between $E_k$ and $E_p$ is disrupted by wave breaking. However, we confirmed this does not affect the analysis of growth and decay rates, since $\mathrm{d}E_W/\mathrm{d}t \approx \mathrm{d}E/\mathrm{d}t$. The evolution equation for $E_W$ is derived from the action balance equation, a general function of the frequency $\omega$ and wave number vector $\mathbf{k}$~\citep{janssen2004interaction}. For a nearly monochromatic wave field evolving over a time scale of $\mathcal{O}(10)T$, the dependence of the wave energy $E_W$ on $\omega$ and on $\mathbf{k}$ can be neglected, as can nonlinear interactions, which do not have sufficient time to develop over such a short duration. Under these assumptions, the evolution of $E_W$ is given by~\citep{peirson2008wind,grare2013growth}
\begin{equation}\label{eqn:amp_t}
  \dfrac{\mathrm{d}E_W}{\mathrm{d}t} = S_{in}-S_d\mathrm{.}
\end{equation}                 
In equation~\eqref{eqn:amp_t}, $S_d$ is the volume-integrated energy dissipation $S_d=\rho_w\int_{\Omega_w} \varepsilon(x,y,z)\,\ dV$ with $\varepsilon(x,y,z)$ given by equation~\eqref{eqn:diss_loc}. \revB{The dissipation term $S_d$ incorporates both viscous effects near the free surface and the energy losses associated with breaking, with the latter being the dominant contribution. Note that the loss due to bottom friction~\citep{grare2013growth} is negligible for deep water waves.} \par
\begin{figure*}[h!]
  \centering
  \includegraphics[trim={0cm 0.0cm 0cm 0cm},width=\textwidth]{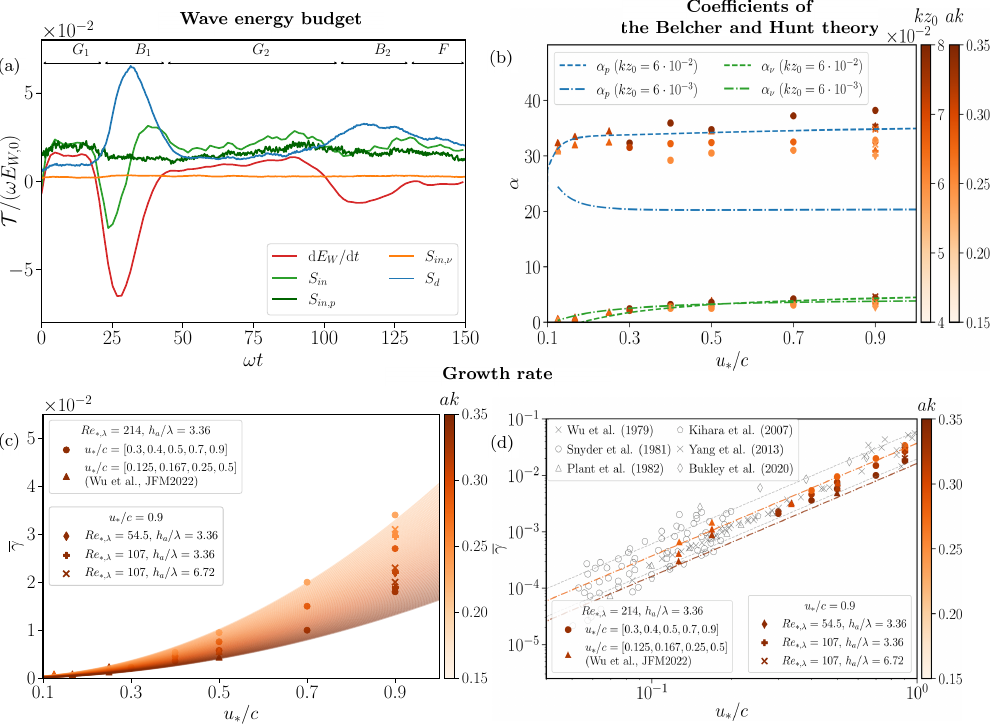}
  \caption{
  (a) Wave energy budget (equation~\eqref{eqn:amp_t}) for $u_\ast/c=0.9$. The $y$-axis shows the different terms in eq. 2, $\mathcal{T}$, being $\mathrm{d}E_W/\mathrm{d}t$, the total wind input $S_{in}$, and dissipation $S_d$, each normalized by $\omega E_{W,0}$ with $E_{W,0}=E_W(0)$. 
  (b) Coefficients $\alpha_{p,\nu}$ from Belcher \& Hunt theory as a function of $u_\ast/c$ and $ak$, with one data point from the first growth stage and two from the second growing stages. The wave slope $ak$ is colorcoded. \revB{The surface roughness $z_0$ is extracted from the mean air velocity profiles, and color-coded.  As a reference, the dashed and dot–dashed lines represent the coefficients $\alpha_{p,\nu}$ computed using a fixed roughness of $kz_0=6\times10^{-2}$ and $kz_0=6\times10^{-3}$, using equations~\eqref{eqn:alpha_p} and~\eqref{eqn:alpha_nu}. The symbols follow the same convention as in panels (c)-(d).} \revB{(c) Non-dimensional growth rate $\overline{\gamma}$ as a function of $u_\ast/c$ computed from the DNS using~\eqref{eqn:gamma_inst_2}}. The shaded area shows estimates from equation~\eqref{eqn:gamma_BH} with $\overline{\alpha}_p=31.4$ and $\overline{\alpha}_\nu=4.6$ for $ak$ spanning $\in[0.15,0.33]$. \revB{Up triangles are present DNS data; circles are DNS from~\citet{wu2022revisiting}}. \revB{In (d) we report $\overline{\gamma}$ as in (c) together with other datasets from Large Eddy Simulations~\citep{yang2013dynamic,kihara2007relationship}, field measurements, and laboratory measurements~\citep {wu1979experimental,snyder1981oceanic,buckley2020surface}. The orange and the dark red dot-dashed lines show theoretical predictions from equation~\eqref{eqn:gamma_BH} using $ak_{\mathrm{min}}=0.15$ and $ak_{\mathrm{max}}=0.33$, respectively. The gray dashed lines are from~\citet{plant1982relationship}.}
  }
  \label{fig:growth}
\end{figure*}

The wind input $S_{in}$ can be estimated directly from the wave energy, equation~\eqref{eqn:amp_t}, as $S_{in}=S_d+\mathrm{d}E_W/\mathrm{d}t$ as proposed in~\cite{donelan2006wave,wu2022revisiting}. Separately, the wind input $S_{in}$ can be computed as the surface-integrated energy fluxes contributing to the wave growth. It can be decomposed into a pressure $S_{in,p}$ and a viscous $S_{in,\nu}$ contribution. Following~\citep{grare2013wave,wu2021wind,wu2022revisiting}, the pressure component can be approximated as the product between the pressure stress and the wave speed, i.e. $S_{in,p}=\tau_{p,x}c$, whereas the viscous component as the product between the viscous stress and the orbital velocity $u_o$, i.e. $S_{in,\nu}=\tau_{\nu,x}u_o$. \revB{Here, the pressure and the viscous stresses are computed as $\tau_{p,x}=-\int_{\Gamma}p\mathbf{n}\cdot\mathbf{e}_xdS$ and $\tau_{\nu,x}=2\rho_a\nu_a\int_\Gamma(\mathbf{D}\mathbf{n})\cdot\mathbf{e}_xdS$, where $p$ is the aerodynamic pressure in the air, $\mathbf{n}$ is the normal vector at the interface, $\mathbf{e}_x=(1,0,0)$ is the unit vector and $\mathbf{D}=(\nabla\mathbf{u}+\nabla\mathbf{u}^T)/2$ is the viscous stress tensor~\citep{scapin2025mom}. From $S_{in}$, an effective or wave-coherent stress $\tau_W$ can be defined as}
\begin{equation}\label{eqn:tau_W}
  \tau_W = \dfrac{S_{in}}{c} = \tau_{p,x}+\tau_{\nu,x}\dfrac{u_o}{c}\mathrm{.}
\end{equation}
Note that equation~\eqref{eqn:tau_W} is based on a linear wave decomposition (using the wave speed and orbital velocity of the wave component), which is valid only during the growing stages. During the breaking stage, we estimate $S_{in}$ using~\eqref{eqn:amp_t}, directly based on the time derivative of the wave energy. The viscous contribution during wave growth is negligible, as shown in figure~\ref{fig:growth}(a). \par
The energy budget terms from equation~\eqref{eqn:amp_t} are shown in figure~\ref{fig:growth}(a) for $u_\ast/c=0.9$, normalized by $\omega E_{W,0}$. The evolution follows three stages: the first growing cycle, the breaking stage, and the second growing cycle. Initially, the accumulation term $\mathrm{d}E_W/\mathrm{d}t$ and the wind input $S_{in}$ dominate, with dissipation mainly due to viscous friction. The wind input term $S_{in}$ closely matches its pressure component $S_{in,p} \approx \tau_{p,x}c$ during this stage, consistent with~\citet{wu2022revisiting}. During the breaking stage, the linear assumptions used to compute $S_{in,p}$ break down, leading to a mismatch between $S_{in}$ and $S_{in,p}$. Here, $S_{in}$ becomes negative as the wave stops absorbing energy from the airflow, and the dissipation term $S_d$ increases sharply, dominating the budget as wave energy is transferred into turbulent kinetic energy in the water column. The viscous contribution $S_{in,\nu}$ remains much smaller than $S_{in,p}$ throughout but shows a slight increase following breaking due to enhanced viscous stress. In the second growing cycle, $S_{in}$ becomes positive again but is lower than in the first cycle due to reduced $ak$ and $\tau_{p,x}$, while $S_{d}$, although reduced compared to the breaking stage, is higher than in the first cycle since the underwater flow remains turbulent. \par 
\subsection{Comparison of the growth rate to existing theoretical framework}\label{sec:growth_rate}
We now aim to quantitatively predict the wave growth rate resulting from wind input. The wind input is used to estimate the instantaneous, non-dimensional wave growth rate, $\gamma$, defined as~\citep{belcher1993turbulent,belcher1999wave}
\begin{equation}\label{eqn:gamma_inst}
  \gamma(t) = \dfrac{S_{in}(t)}{\omega E_W(t)}\mathrm{.}
\end{equation}
In practice, $\gamma(t)$ will be evaluated in an averaged sense over given time windows delimited by $t_1$ and $t_2$: 
\begin{equation}\label{eqn:gamma_inst_2}
  \overline{\gamma} = \dfrac{1}{t_2-t_1}\int_{t_1}^{t_2} \dfrac{S_{in}(t)}{\omega E_W(t)}dt\mathrm{.}
\end{equation}
We will consider $\overline{\gamma}$ evaluated over the growing stages $G_1$ and $G_2$, which can then be compared with theoretical frameworks that describe its dependence on wave parameters, particularly the wave slope $\overline{ak}$ and the normalized friction velocity $u_\ast/c$. \par
Following~\cite{peirson2008wind,fedorov1998nonlinear,melville2015equilibrium,buckley2020surface}, and using $S_{in}=c\overline{\tau}_W$, $E_W=\rho_wga^2/2$ and the dispersion relation $c=\sqrt{g/k}$, we obtain
\begin{equation}\label{eqn:gamma_00}
  \overline{\gamma}_M = \dfrac{2\overline{\tau}_W}{\rho_w c^2\overline{ak}^2}\mathrm{.}
\end{equation}
\revB{In equation~\eqref{eqn:gamma_00}, the time-averaged growth rate $\overline{\gamma}$ for a given wind-wave condition is obtained from the time-averaged wave-coherent momentum flux $\overline{\tau}_W$ (itself given by equation~\eqref{eqn:tau_W}).} Earlier theories summarized in~\cite{kihara2007relationship,wu2022revisiting} based on sheltering~\citep{jeffreys1925formation}, and critical layer~\cite{miles1957generation} lead to $\overline{\tau}_W=\alpha_p(\overline{ak}^2/2) \rho_a u_\ast^2$, with $\alpha_p$ function of the wind structure and wave slope, or assumed to be a constant, leading to the classic form for $\overline{\gamma}$
\begin{equation}\label{eqn:gamma_0}
  \overline{\gamma}_J = \dfrac{\rho_a}{\rho_w}\alpha_p\left(\dfrac{u_\ast}{c}\right)^2\mathrm{.}
\end{equation}
\nsc{In the present configuration, for high values of $u_\ast/c$ and steep waves, the critical layer, i.e. the height where $\langle u \rangle(z) = c$, lies within the viscous sublayer, where the assumptions behind Miles' original theory, particularly the inviscid and quasi-laminar approximations, no longer hold. While later extensions~\citep {miles1959generation} attempt to incorporate viscous effects, they still rely on the quasi-laminar assumption and neglect wave-induced turbulence, which becomes important for steep waves. \citet{belcher1993turbulent} relaxes these assumptions and accounts for pressure perturbations arising from turbulent sheltering, i.e. asymmetric distortion of turbulent eddies by the wave, which remains effective even when the critical height lies within the roughness or viscous sublayer. \par
\citet{belcher1993turbulent,belcher1999wave} compute $\overline{\tau}_W$ under such assumptions}. They separate the turbulent boundary layer into an inner and an outer region. In the inner region, close to the wave surface, turbulent eddies are assumed in equilibrium with the mean velocity gradient. The characteristic timescale for an eddy to decorrelate and interact with another one is $T_L=\kappa h/u_\ast$ \revB{(with $\kappa$ the von Kármán constant and $h$ the undisturbed reference height above the water surface)} and turbulence is described with a mixing-length model~\citep{townsend1951structure}. In the outer region, eddies are rapidly distorted by the flow with a timescale $T_A=k/[\overline{u}(h)-c]$, and described using the rapid distortion theory~\citep{belcher1993turbulent}. Assuming negligible flow separation (i.e. non-separated sheltering) over the wave surface,\revB{~\cite{belcher1993turbulent} employed} asymptotic expansions to evaluate the surface pressure and viscous stress distribution, leading to $\overline{\tau}_{W,p}$ and $\overline{\tau}_{W,\nu}$ as
\begin{equation}\label{eqn:tau_Wp}
  \overline{\tau}_{W,p}=\dfrac{\alpha_p\overline{ak}^2/2}{1+\alpha_p\overline{ak}^2/2}\rho_au_\ast^2\mathrm{,}
\end{equation}
\begin{equation}\label{eqn:tau_Wnu}
  \overline{\tau}_{W,\nu}=\dfrac{\alpha_\nu\overline{ak}^2/2}{1+\alpha_p\overline{ak}^2/2}\rho_au_\ast^2\mathrm{,}
\end{equation}
where the terms $\alpha_{p,\nu}$ are the pre-factors of the pressure and viscous stress momentum fluxes that can be explicitly calculated with knowledge of the wind profile. By substituting equations~\eqref{eqn:tau_Wp} and~\eqref{eqn:tau_Wnu} into~\eqref{eqn:gamma_0}, we obtain
\begin{equation}\label{eqn:gamma_BH}
  \overline{\gamma}_{BH} = \dfrac{\rho_a}{\rho_w}\dfrac{2(\alpha_p+\alpha_\nu)}{2+\alpha_p\overline{ak}^2}\left(\dfrac{u_\ast}{c}\right)^2\mathrm{.}
\end{equation}
Note that equation~\eqref{eqn:gamma_BH} contains the explicit dependence on the wave steepness $\overline{ak}$. For negligible viscous stress work, i.e. $\alpha_\nu\ll\alpha_p$, and for 'small' $\overline{ak}$, i.e. $\overline{ak}\lesssim 0.15$, $\overline{\tau}_{W,p}\approx \alpha_p(\overline{ak}^2/2)\rho_a u_\ast^2$, equation~\eqref{eqn:gamma_BH} reduces to \revB{equation~\eqref{eqn:gamma_0}} from~\cite{miles1957generation,jeffreys1925formation}. \par
The framework developed in~\cite{belcher1999wave} provides a closed expression for $\alpha_{p,\nu}$, which read as follows
\begin{equation}\label{eqn:alpha_p}
  \alpha_p = 2\left(\dfrac{\langle u\rangle_m-c}{\langle u\rangle_i-c}\right)^4\left(2-\dfrac{c}{\langle u\rangle_i}\right)-2\left(\dfrac{\langle u\rangle_m-c}{\langle u\rangle_i-c}\right)^2\dfrac{c}{\langle u\rangle_i}+2\kappa\delta^{2n}\left(\dfrac{\langle u\rangle_m-c}{u_\ast}\right)\mathrm{,}
\end{equation}
\begin{equation}\label{eqn:alpha_nu}
  \alpha_\nu = \dfrac{2(\langle u\rangle_m-c)^2}{(\langle u\rangle_i-c)u_i}-\dfrac{2c}{\langle u\rangle_i}\mathrm{.}
\end{equation}
In equations~\eqref{eqn:alpha_p} and~\eqref{eqn:alpha_nu}, the velocities $\langle u\rangle_i$ and $\langle u\rangle_m$ are the streamwise velocity evaluated at two boundary-layer heights, $z_i$ and $z_m$. The height $z_i$ is where $T_A = T_L$, and is computed iteratively from $kz_i|\log(z_i/z_0)-\kappa c/u_\ast|=2\kappa^2$. The mid-layer height is defined as $kz_m=\sqrt{\delta/\kappa}$ with $\delta=u_\ast/|u(\lambda/2\pi)|$~\citep{belcher1993turbulent}. These expressions reduce the calculation of $\overline{\tau}_W$ and $\overline{\gamma}$ to evaluating the velocity profile at three heights: $z=[z_i,z_m,\lambda/2\pi]$. 
Figure~\ref{fig:growth}(b) shows the pre-factors $\alpha_{p,\nu}$ directly extracted from the DNS velocity profiles (shown and discussed in~\cite{scapin2025mom}), for different values of $u_\ast/c$. Data point colors indicate different instantaneous $\overline{ak}$ values from the first growing cycle, together with the corresponding surface roughness $z_0$, and from two time windows of equal size in the second growing cycle. Note that because we have a narrow-banded idealized configuration, we observe a one-to-one relationship between $z_0$ and $ak$ (which is not general to a broad-banded wave field).
For $u_\ast/c \in [0.125, 0.9]$, DNS results show $\alpha_{p,\nu}$ is relatively insensitive to $u_\ast/c$ but depends on surface roughness $z_0$ (evaluated by considering a log-layer for the mean profile, $\langle u\rangle=(u_\ast/\kappa)\log(z/z_0)$), decreasing as $kz_0$ (and $ak$) decrease. However, since $kz_0$ varies narrowly ($[4\cdot 10^{-2}, 8\cdot 10^{-2}]$), this variation is moderate. Note that the pressure contribution $\alpha_p$ is consistently much larger than the viscous contribution $\alpha_\nu$, with $\alpha_\nu/\alpha_p$ increasing from $0.04$ to $0.12$ with $u_\ast/c$. Given this limited variability in $kz_0$, we adopt mean values $\overline{\alpha}_p = 31.4$ and $\overline{\alpha}_\nu = 4.6$, used in equation~\eqref{eqn:gamma_BH} to compute growth rates over $ak \in [0.15, 0.35]$. \par
Figure~\ref{fig:growth}(c) clearly shows the modulation of $\overline{\gamma}$ induced by wave steepness for a fixed $u_\ast/c$ and emphasizes the importance of its inclusion in the expression of wave-coherent stresses and the growth rate. We observe in our DNS with strong wind and steep waves that at a fixed $u_\ast/c$, the non-dimensional growth decreases with increasing $ak$, a feature that is very well captured by Belcher's theory.
Figure~\ref{fig:growth}(d) combines the estimation of $\overline{\gamma}$ from our DNS, the ones retrieved from~\citep{wu2022revisiting} at lower $u_\ast/c$ and a collection of numerical and experimental data~\citep{wu1979experimental,snyder1981array,plant1982relationship,kihara2007relationship,yang2013dynamic,buckley2020surface}. Furthermore, we include the estimation of the theoretical growth rate following~\eqref{eqn:gamma_BH} (black dot-dashed) with $\overline{\alpha}_p=31.4$, $\overline{\alpha}_\nu=4.6$ for $ak_\mathrm{min}=0.15$, $ak_\mathrm{max}=0.35$, and the one obtained in~\cite{plant1982relationship} (grey dashed lines), who suggested $\overline{\gamma}=(0.04\pm 0.02)(u_\ast/c)^2$ (equation~\eqref{eqn:gamma_0}). \par
As discussed in previous studies, the values of $\overline{\gamma}$ at a given $u_\ast/c$ exhibit a large scatter, reaching up to an order of magnitude. This arises from several factors. Estimating the wind input $\overline{S}_{in}$ is challenging due to uncertainties in measuring aerodynamic pressure~\citep{grare2013growth}, as well as the dependence on the wind velocity profile and the lack of direct measurements. Reported coefficients fitted from observed growth rates, using equation~\eqref{eqn:gamma_0} as in~\cite{peirson2008wind} can vary by an order of magnitude~\citep{belcher1999wave}. The framework of~\citet{belcher1993turbulent} provides a physical approach to evaluate $\alpha_{p,\nu}$ based on the wind profile and highlights the wave slope modulation as a source of scatter in the growth rate at a given $u_\ast/c$. In the high $u_\ast/c$ and $ak$ regime, the critical height becomes very close to the surface layer, and a sheltering approach such as the one \revA{from~\cite{belcher1999wave}}, incorporating the effect of turbulence, becomes preferable.
%
%
%
%
\section{Wave breaking-induced dissipation}\label{sec:diss}
In this section, we discuss the turbulence dissipation induced by wave breaking. In section~\ref{sec:b_param}, we confirm that the dissipation caused by wave breaking is controlled by the wave slope at breaking and is quantified by the inertial scaling originally proposed by \cite{drazen2008inertial}. In section~\ref{sec:prof}, we analyze the vertical profiles at different stages of the wave field and demonstrate that wave breaking promotes a transition to turbulence, with $\langle\varepsilon\rangle(z)\propto z^{-1}$, and that the inertial scaling for the dissipation due to breaking can be used to unify the dissipation profiles at different $u_\ast/c$. 
\begin{figure*}[h!]
  \centering
  \includegraphics[trim={0cm 0.0cm 0cm 0.75cm},width=\textwidth]{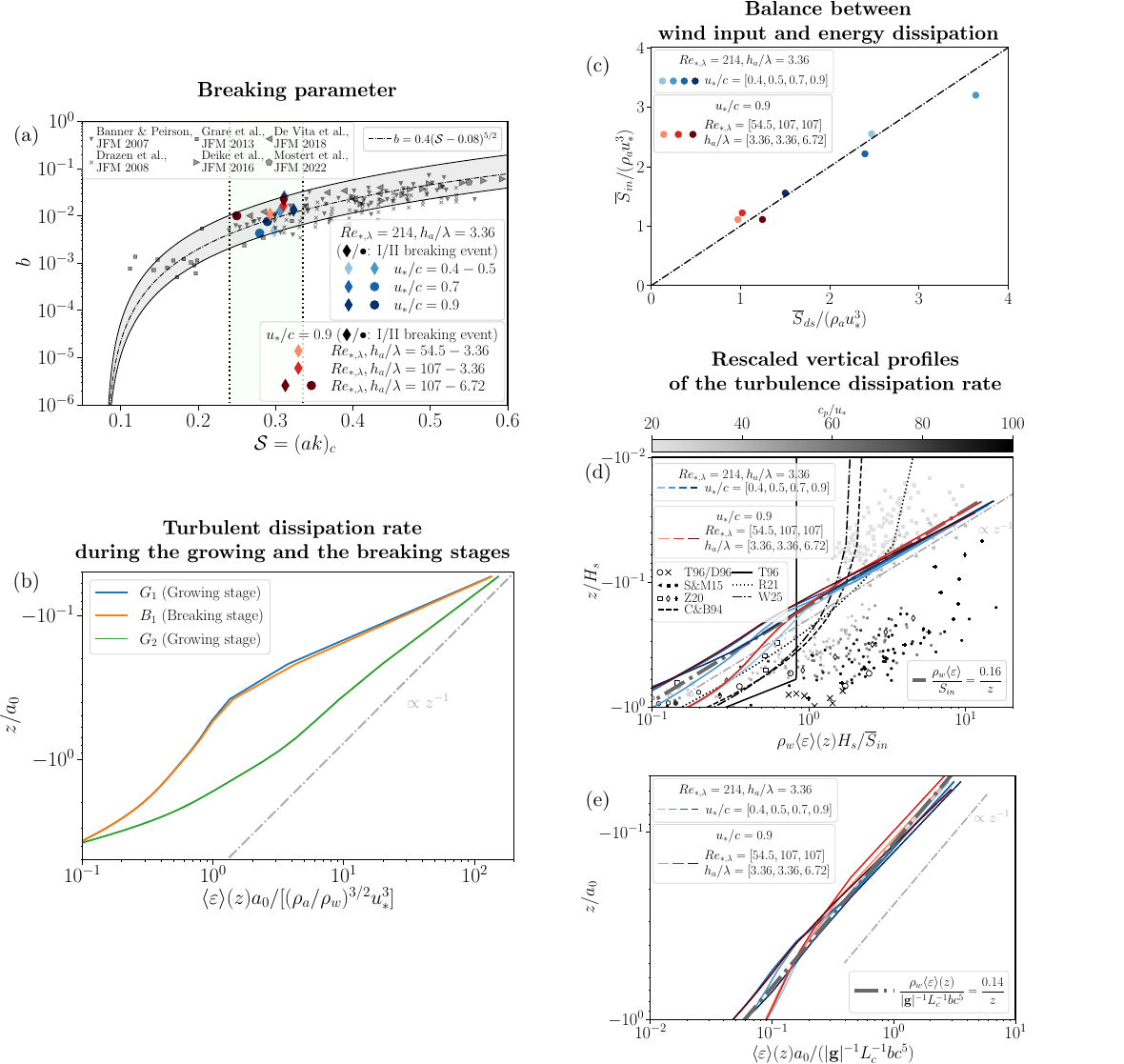}
  \caption{
  (a) Breaking parameter $b$ versus wave steepness for various $u_\ast/c$. Solid lines show equation~\eqref{eqn:b_param} with $\chi_0=0.25$–$0.95$; vertical dot-dashed lines indicate the range of the critical steepness at breaking $(ak)_c=0.28$–$0.33$; black dot-dashed line is $b=0.4(\mathcal{S}-0.08)^{5/2}$~\citep{deike2016air}. \revB{For $u_\ast/c=0.7-0.9$, $b$ is shown for the first ($\blacklozenge$ diamond) and second breaking events ($\bullet$ circle)}. Experimental data with/without wind~\citep{banner2007wave,drazen2008inertial,grare2013growth} and DNS without wind~\citep{deike2016air,de2018breaking,mostert2022high} are also included.
  (b) Time-averaged dissipation profiles for $u_\ast/c=0.9$ during $G_1$, $B_1$, and $G_2$, normalized by $(\rho_a/\rho_w)^{3/2}u_\ast^3/a_0$.
  (c) Normalized wind input $\overline{S}_{in}=\overline{\tau}_{W,p}c$ over $G_1$ versus breaking-induced dissipation $S_{ds}=\rho_wg^{-1}L_c^{-1}b c^5$ over $B_1$. 
  (d) \revB{Dissipation profiles after breaking ($G_2$), normalized by significant wave height $H_s$ and wind input, $ \rho_w\langle \varepsilon\rangle(z) H_s/\overline{S}_{in}$. DNS data are in colored lines for different parameters ($u_\ast/c$, Reynolds number). All profiles collapse on a single curve following $z^{-1}$. DNS profiles are compared with field observations (grey, black symbols) from~\citep{terray1996estimates,drennan1996oceanic,sutherland2015field,zippel2020measurements,thomson2012wave}, where $\overline{S}_{in}$ is estimated from measurement and parameterization of wind input. The best fit lines of the dissipation $\ langle \ varepsilon (z)\rangle$ as a function of the water depth proposed in~\citep{craig1994modeling,terray1996estimates,romero2021representing,wu2025turbulence} are also reported (dashed, solid, dotted, and dashdot lines).}
  (e) Depth normalized by $a_0$ and dissipation normalized by $\overline{S}_{ds}/(\rho_w a_0)$.
  }
  \label{fig:diss}
\end{figure*}
\subsection{Estimation of the breaking parameter}\label{sec:b_param}
We characterize the energy loss during the wind-forced breaking events and compare with previous results obtained in the absence of wind. We demonstrate the applicability of the inertial scaling law for plunging and spilling breakers~\citep{drazen2008inertial,pizzo2013vortex}, which adapted the classical turbulent dissipation scaling~\citep{taylor1938spectrum,vassilicos2015dissipation} to wave breaking. \par
Following~\citet{drazen2008inertial}, the local dissipation rate is estimated using Taylor's frozen turbulence hypothesis, $\varepsilon=C_\varepsilon w_c^3/h_c$, with $C_\varepsilon=\mathcal{O}(1)$, $h_c=a$ the breaking height, and $w_c=\sqrt{2ga}$ the ballistic velocity. The breaking-induced turbulence is assumed confined to a volume $AL_c$, where $A\approx \pi a^2/4$~\citep{duncan1981experimental,drazen2008inertial} and $L_c$ is the crest length ($L_c=4\lambda$ in our setup). 
The dissipation per unit crest length is $\varepsilon_l = \rho_w \varepsilon A / L_c$~\citep{duncan1981experimental,phillips1985spectral,deike2022mass}, which leads to $\varepsilon_l = b\rho_wc^5/g$, with $b = \chi_0\mathcal{S}^{5/2}$, $\chi_0=(\pi/\sqrt{2})C_\varepsilon$, and $\mathcal{S}=(ak)_c$ the breaking slope. Introducing a threshold slope $\mathcal{S}_0\approx 0.08$~\citep{romero2012spectral,grare2013growth}, we have
\begin{equation}\label{eqn:b_param}
  b=\chi_0(\mathcal{S}-\mathcal{S}_0)^{5/2}\mathrm{.}
\end{equation}
\revB{To calculate the breaking parameter $b$,} we consider the temporal variation of the wave energy $E_W=2E_p$ as \revB{reported in figure~\ref{fig:fig1}(c)} for $u_\ast/c=0.9$. We define two physical times, $t_i$ and $t_f$, as the start and end of breaking, and evaluate $\varepsilon_l=2(E_{p,i}-E_{p,f})/(L_c(t_i-t_f))$, where $E_{p,i}(t=t_i)$ and $E_{p,f}(t=t_f)$ are defined as $E_{p,i}=\max(E_p)$ and $E_{p,f}=\min(E_p)$ over a single growing/breaking cycle. \par 
In figure~\ref{fig:diss}(a), we compare breaking parameters from our DNS with laboratory experiments on wave breaking~\citep{banner2007wave,drazen2008inertial} and the scaling $b\sim \mathcal{S}^{5/2}$  with a range of the scaling pre-factor $\chi_0$, (of order $\mathcal{O}(1)$) from $\chi_0 \in [0.25-0.95]$, as suggested in the literature, by experimental and numerical data~\citep{romero2012spectral,deike2015capillary,mostert2022high,deike2016air}. The DNS data of wind-forced breaking waves at high wind speed show good agreement with both the theoretical scaling and laboratory experiments across the full range of tested $u_\ast/c$. \revA{Wind forcing slightly alters the breaking threshold, ranging from $(ak)_c = 0.31$ at $u_\ast/c = 0.4$ to $(ak)_c = [0.28-0.34]$ at $u_\ast/c = 0.9$. Consequently, each case exhibits a different level of dissipation, as the variation in $(ak)_c$ leads to a corresponding change in $b$ according to the inertial scaling argument~\citep{drazen2008inertial}.} \par
We conclude that the scaling $b\propto \mathcal{S}^{5/2}$, originally developed for breaking waves without wind input~\citep{kendall1994energy,drazen2008inertial} applies in the context of wind-forced breaking waves. This suggests that breaking is a universal process, with its decay rate largely independent of the specific cause triggering the event, such as wind input, and primarily dependent on the amplitude at breaking. Using the breaking height as characteristic length and the characteristic ballistic velocity, the breaking-induced dissipation is determined by a single parameter $b$ for breaking waves due to linear and non-linear focusing, as well as wind-forcing~\citep{iafrati2009numerical,deike2015capillary,de2018breaking,deike2016air,mostert2022high,grare2013growth,drazen2008inertial,kendall1994energy,banner2007wave}.
\subsection{Vertical profiles of the dissipation}\label{sec:prof}
We now examine the vertical distribution of turbulent dissipation. The local dissipation rate (equation~\eqref{eqn:diss_loc}), is averaged over the two periodic directions, yielding the mean profile as a function of depth $\langle \varepsilon \rangle(z)$. Following~\cite{wu2022revisiting,scapin2025mom}, profiles are plotted in a wave-following coordinate system to ensure that the averaging accounts for both wave crests and troughs. 
\par 
Figure~\ref{fig:diss}(b) shows the dissipation profiles normalized by $(\rho_a/\rho_w)^{3/2}u_\ast^3/a_0$ as a function of water depth $z/a_0$, for $u_\ast/c = 0.9$. The three profiles are time-averaged over the first growing and breaking stages ($G_1$-$B_1$), and the second growing cycle ($G_2$). The dissipation profile steepens and approaches a $z^{-1}$ scaling during $B_1$, with high dissipation near the surface. \revB{In the post-breaking stage $G_2$, regions of high dissipation extend to around $\approx 2a_0$ (also visible in figure~\ref{fig:fig1}(g)), a rapid transition due to breaking. Below this depth, the magnitude of turbulence dissipation becomes negligible and rapidly decays to zero}. \par
Under strongly forced wind conditions, characterization of the wave spectra, breaking distribution and associated upper ocean turbulence have often invoked the assumption of equilibrium between the mean energy input from the wind $\overline{S}_{in}$ and the total mean dissipation due to breaking $\overline{S}_{ds}$~\citep{phillips1985spectral,drennan1996oceanic,terray1996estimates,sutherland2015field,zippel2020measurements,wu2025turbulence,zippel2022parsing,wu2023breaking}. \par
\revB{We directly assess the balance between wind input (evaluated during the growth stage) and energy dissipation (evaluated during the breaking stage) in the present set-up for an individual breaking event. This balance should be interpreted as the wind energy injected during growth being dissipated during the subsequent breaking.} Figure~\ref{fig:diss}(c) shows the wind input $\overline{S}_{in}$ extracted directly from stress resolved in the DNS (during the growth stage) as a function of the total dissipation $\overline{S}_{ds}$, evaluated directly from the turbulence resolved in the DNS (during the breaking stage) as $\overline{S}_{ds} = \rho_wg^{-1} \int b \Lambda(c) c^5 dc$ (the fifth moment of the breaking distribution $\Lambda(c)$, following~\cite{duncan1981experimental,phillips1985spectral}), which simplifies to $\overline{S}_{ds}=\rho_wg^{-1} L_c^{-1}bc^5$ in our narrowbanded wave conditions. The balance between the two terms ($\overline{S}_{in}$ and $\overline{S}_{ds}$) occurring subsequently in time is well observed. \par
To compare results across different wind conditions,~\citet{terray1996estimates,drennan1996oceanic,sutherland2015field} presented the vertical profiles of turbulence dissipation rate $\langle \varepsilon \rangle (z)$ normalized by the wind input $\overline{S}_{in}$, using the significant wave height $H_s$ as characteristic vertical height. This normalization is shown in figure~\ref{fig:diss}(d), together with field data from~\citet{sutherland2015field}, \revA{and refers to the second growth cycle $G_2$, when the underwater flow has transitioned to turbulence.} The numerical profiles of $\rho_w \langle \varepsilon \rangle (z) H_s / \overline{S}_{in}$ exhibit an excellent collapse up to $0.1 H_s$, maintaining a $z^{-1}$ scaling and are well described by 
\begin{equation}\label{eqn:verteps_s}
  \langle\varepsilon\rangle(z) = A\dfrac{\overline{S}_{in}/\rho_w}{z}\mathcal{,}
\end{equation}
where $A \approx 0.16$ is a non-dimensional constant obtained by best fit to the data.
\revB{The collapsed numerical data fall within the scatter of field observations reported in previous studies~\citep{craig1994modeling,terray1996estimates,drennan1996oceanic,sutherland2015field,zippel2020measurements}, though some differences with field data inevitably arise.}
\revB{The scatter in field measurements reflects uncertainties in dissipation estimates and wind input, with reported error bars in the different studies spanning up to an order of magnitude. Moreover, estimates in the field rely on time-averaged wind input and dissipation over the same intervals, without distinguishing between growing and breaking stages. In contrast, in the DNS, we directly compute wind input from the resolved stress field and dissipation from underwater turbulence, over time windows restricted to the relevant processes: the growing stage for wind input and the breaking stage for dissipation.}
\revB{Some field observations report a constant-dissipation layer beneath the wave field~\citep{drennan1996oceanic,terray1996estimates,sutherland2015field}, which, as noted by~\citet{wu2025turbulence}, arises from using an absolute reference frame, whereas we adopt a wave-following one. Our dissipation profiles are obtained from the wave-following coordinate defined on a single, well-resolved growing and breaking cycle. In field conditions, dissipation is averaged over many events (growing and breaking) of different scales and phases, so that some variability in $\langle\varepsilon\rangle(z)$ might be related to the superposition of events rather than different mechanisms.
Finally, the shallower $z^{-1}$ layer in our DNS compared to field data is due to the single breaking event, while field observations and recent modeling capture the cumulative effect of multiple events, deepening the turbulence layer.}

\revB{An immediate implication of the scaling in equation~\eqref{eqn:verteps_s} is that it is inconsistent with the wall-layer scaling for dissipation, $\langle\varepsilon\rangle_{wl}(z) = u_{\ast,w}^3/(\kappa z)$, where $u_{\ast,w} = u_\ast\sqrt{\rho_a/\rho_w}$ is the water-side friction velocity and $\kappa = 0.41$ is the von Kármán constant, which has sometimes been invoked as a default scaling for $\langle\varepsilon\rangle(z)$.} Indeed, equation~\eqref{eqn:verteps_s} implies $\langle\varepsilon\rangle(z) \sim u_\ast^2 c / z$ since $\overline{S}_{in}\sim u_\ast^2 c$, while the wall-layer scaling suggests $\langle\varepsilon\rangle(z) \sim u_\ast^3 / z$. Therefore, the physical argument behind the wall-layer scaling (based on shear-induced turbulence) is not appropriate for characterizing the dissipation profiles below breaking waves at different values of $u_\ast/c$ (see the supplementary material). \par
Since we have observed the balance $\overline{S}_{in} \approx \overline{S}_{ds}$ (figure~\ref{fig:diss}(c)), we expect that normalizing $\langle\varepsilon\rangle(z)$ by $\overline{S}_{ds}$ will unify the profiles across different $u_\ast/c$. The results are shown in figure~\ref{fig:diss}(e), where we observe a good collapse of $\langle\varepsilon\rangle(z)$ across $u_\ast/c$. The vertical turbulence dissipation profiles are well described by 
\begin{equation}\label{eqn:eps_sdiss} 
  \langle\varepsilon\rangle(z) = \mathcal{A} \dfrac{\overline{S}_{ds}/\rho_w}{z} = \mathcal{A} \dfrac{g^{-1} L_c^{-1} b c^5}{z}\mathcal{.} 
\end{equation}
with $\mathcal{A}\approx 0.14$ similar to $A$ in equation~\eqref{eqn:verteps_s}. The underwater dissipation is thus fully described by the breaking parameter $b$ (linked to the wave slope at breaking) and the wave speed $c$, consistent with the inertial dissipation \cite{drazen2008inertial} and the original breaking dissipation from~\cite{duncan1981experimental,phillips1985spectral}. \par
We highlight that the observed scaling of dissipation profiles with wind input (equation~\eqref{eqn:verteps_s}, figure~\ref{fig:diss}(d)), and with breaking-induced dissipation (equation~\eqref{eqn:eps_sdiss}, figure~\ref{fig:diss}(e)), aligns with our understanding of wind-wave growth (via sheltering) and dissipation by breaking (via inertial scaling) combined with the idea of balance between wind input and breaking. Wave breaking occurs when fluid inertia overcomes restoring forces such as gravity and surface tension. After the onset of breaking, wind input no longer affects the dissipation rate; however, the total wind energy transferred to the waves remains crucial in determining the breaking conditions. Therefore, underwater turbulence dissipation naturally scales both with the total wind input (equation~\eqref{eqn:verteps_s}) and local steepness at breaking (equation~\eqref{eqn:eps_sdiss}) under local equilibrium conditions.
%
%
%
%
\section{Conclusions}\label{sec:concl}
We performed direct numerical simulations of wind-forced breaking waves in a high-wind-speed regime, capturing the wind-driven growth of a narrowband wave field up to breaking, followed by energy dissipation and the resulting underwater turbulence structure. Building on our previous work~\citep{scapin2025mom}, we divide the evolution into growth and breaking stages, and analyze each separately. \revB{This approach provides insights into small-scale turbulence processes associated with breaking waves at high wind speed, and highlights the splitting of energy fluxes during wave growth, where wind energy is transferred to the waves, and during breaking, where energy is transferred into turbulence in the upper ocean, and the relative balance between these subsequent processes.} \par 
First, we confirm that during the growing stage, wave growth is primarily driven by pressure drag. During the breaking stage, the energy loss from the wave field becomes the dominant mechanism of energy transfer to the water column, further accelerating the development of underwater drift and promoting the transition to turbulence in the subsurface currents. This, in turn, enhances turbulence dissipation in the underwater flow. We quantify the wave growth rate and demonstrate that it is strongly modulated by the wave steepness, which contributes to the variability in growth rate estimates at the same $u_\ast/c$ observed in the literature. \par
During wave breaking, we measure the energy loss and demonstrate that it follows the inertial scaling law $b \sim \mathcal{S}^{5/2}$, consistent with prior experiments and simulations conducted in the absence of wind forcing. This confirms that the energy loss is primarily governed by the wave amplitude at the onset of breaking and is insensitive to wind input. \par
Finally, we examine energy dissipation and show that, following breaking events, vertical dissipation profiles exhibit a clear $z^{-1}$ scaling near the surface, in agreement with field observations. When normalized by the wind input $\overline{S}_{in}$, these profiles collapse across a wide range of $u_\ast/c$, supporting the idea that $\rho_w\langle\varepsilon\rangle(z) H_s/\overline{S}_{in}$ is an appropriate scaling to characterize near-surface turbulence generated by breaking waves. Moreover, we observe a balance between wind input (evaluated during growth) and energy dissipation (during breaking), confirming the assumption that the total wind energy input is transferred to the water column via breaking. Building on this, we propose a rescaling of the dissipation profiles using the fifth moment of the breaking distribution, i.e. $\langle\varepsilon\rangle(z) = \mathcal{A} (g^{-1} L_c^{-1} b c^5)/z$, which explicitly includes the breaking parameter $b$ and wave speed $c$. This new scaling law also yields an excellent collapse of the dissipation profiles across different values of $u_\ast/c$. \par
Overall, our results confirm that wave breaking, arising when fluid inertia overcomes restoring forces, is a universal process whose energy loss and turbulence generation are largely independent of the specific mechanism that triggers the event. These findings offer new insights into the physical processes governing air-sea energy exchange under strong wind forcing, \revB{with the event splitting analysis possibly providing a new path for analysis of field data and implications for the development of physics-based parameterizations of momentum and energy fluxes in high-wind-speed regimes} relevant to tropical cyclones and winter storms.
\section*{Open Research Statement}\notag 
The open-source Basilisk solver~\citep{Popinet_Basilisk_2014} used in this study is available at \url{https://basilisk.fr/}. The files for the precursor simulations and the files and post-processing scripts for the two-phase simulations are archived at permanent Zenodo links~\cite{Scapin_precursor_2025} and~\cite{Scapin_turbulentSetup_2025}, respectively. Their corresponding GitHub mirrors are available at \url{https://github.com/DeikeLab/precursor} and \url{https://github.com/DeikeLab/turbulent_setup}.
\acknowledgments
This work is supported by the \textit{National Science Foundation} under grant 2318816 to LD (Physical Oceanography program), the \textit{NASA Ocean Vector Winds Science Team}, grant 80NSSC23K0983 to LD and JTF. NS was supported by the fellowship \textit{High Meadows Environmental Institute Postdoctoral Teaching Program}. Computations were performed using the \textit{Stellar} machine, granted by the \textit{Cooperative Institute for Earth System modeling} (CIMES), and managed by Princeton Research Computing. This includes the Princeton Institute for Computational Science and Engineering, the Office of Information Technology's High-Performance Computing Center, and the Visualization Laboratory at Princeton University. NS acknowledges the discussion with Alexander Babanin during the WISE Meeting 2025.
%
%
\bibliography{agubib}
\end{document}


\maketitle
\thispagestyle{firstpage}

\linenumbers
\onehalfspacing
\setstretch{0.95}          
\setlength{\parskip}{1em}  
\section{Methodology: Direct Numerical Simulations}\label{app:method}
\begin{figure*}[h]
  \centering
  \includegraphics[width=0.65\textwidth]{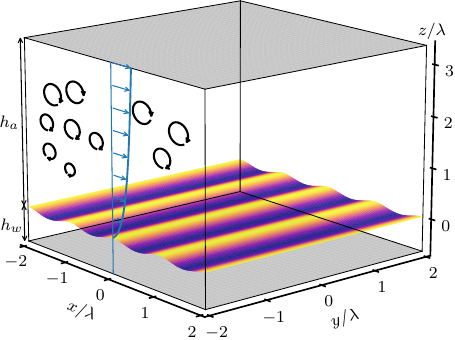}
  \caption{Illustration of the system composed of turbulent airflow ($0\leq z\leq h_a$), the wave field and the water field ($-h_w\leq z\leq 0$). Here, $h_a$ and $h_w$ are the mean air and water mean heights with $\lambda$ the fundamental wavelength. The airflow is a fully developed turbulent boundary layer (mean profile in light-blue line, while turbulent eddies are illustrated in black). In the surface contour, dark-purple regions denote wave troughs, while yellow regions indicate wave crests.}
  \label{fig:model}
\end{figure*}

In this section of the supplementary material, we describe the numerical methodology employed to study wind-forced breaking waves in the high wind-speed regime. The main feature of our approach is to treat the system composed of wind, waves, and water as a two-phase system, i.e., air and water, where the wave field is located at the interface between the two phases, solving for the most general two-phase Navier-Stokes equations without simplifications or subgrid-scale model. A sketch of this configuration is shown in figure~\ref{fig:model}. \par
To distinguish the two phases, an indicator function $\mathcal{H}$ is introduced. It is set equal to $0$ in the air and to $1$ in the water, with a small transition region, i.e. $0<\mathcal{H}<1$, corresponding to the wave interface. The evolution of $\mathcal{H}$ and the wave field is governed by the advection equation
\begin{equation}\label{eqn:ind_func}
\dfrac{\partial \mathcal{H}}{\partial t} + \mathbf{u}\cdot\nabla\mathcal{H} = 0\mathrm{,}
\end{equation}
where $\mathbf{u}=(u,v,w)$ is the three-dimensional velocity field, obtained by solving the incompressible Navier–Stokes equations in both the air and water phases, including surface tension effects. These equations read~\citep{tryggvason2011direct}:
\begin{equation}\label{eqn:cont}
  \nabla\cdot\mathbf{u} = 0\mathrm{,}
\end{equation}
\begin{equation}\label{eqn:momc}
  \dfrac{\partial(\rho\mathbf{u})}{\partial t} + \nabla\cdot(\rho\mathbf{u}\mathbf{u}) = -\nabla p + \nabla\cdot[\mu(\nabla\mathbf{u}+\nabla\mathbf{u}^T)] + \sigma\varkappa\delta_\Gamma\mathbf{n}_\Gamma+\rho g\mathbf{e}_z\mathbf{,}
\end{equation}
where $p$ is the pressure, $\sigma$ is the surface tension coefficient, $g$ is gravity with $\mathbf{e}_z=(0,0,-1)$ is a unit vector, $\varkappa$ is the curvature of the interface, $\mathbf{n}_\Gamma$ is the outward-pointing unit normal to the interface, i.e. the wave field, and $\delta_\Gamma$ is the Dirac distribution which is not zero only at the interface. In equation~\eqref{eqn:momc}, $\rho$ and $\mu$ are the density and dynamic viscosity fields, which are computed using an arithmetic average based on the indicator function as
\begin{equation}\label{eqn:dens}
  \rho = \rho_w\mathcal{H}+\rho_a(1-\mathcal{H})\mathrm{,}
\end{equation}
\begin{equation}\label{eqn:visc}
  \mu = \mu_w\mathcal{H}+\mu_a(1-\mathcal{H})\mathrm{,}
\end{equation}
where subscripts $w$ and $a$ denote the water and air properties, respectively. \par
Equations~\eqref{eqn:ind_func},~\eqref{eqn:cont}, and~\eqref{eqn:momc} are solved using state-of-the-art numerical methods for multiphase flow implemented in the open-source Basilisk solver\footnote{\href{http://basilisk.fr/}{http://basilisk.fr/}}; see e.g.~\citep{popinet2015quadtree,popinet2018numerical,van2018towards} for more details. In particular, equation~\eqref{eqn:ind_func} is solved using a conservative and diffusion-free geometric volume-of-fluid method, which enables accurate reconstruction and advection of the interface between the air and water phases. Equations~\eqref{eqn:cont} and~\eqref{eqn:momc} are solved using a pressure-correction scheme: the velocity is first predicted without including the pressure gradient, and then corrected to enforce the divergence-free condition in equation~\eqref{eqn:cont}. Furthermore, the momentum equation is discretized using a momentum-conserving scheme, which ensures numerical stability even in the presence of high-density ratios, such as those encountered in air–water systems. The entire set of equations is solved on an adaptive grid, providing high resolution near the wave field while maintaining a coarser mesh in the bulk of each phase. \par
The key advantage of this approach is that it does not require any model for wave motion or subgrid-scale turbulence, thereby enabling a realistic, first-principles description of the coupling among wind, the wave field, and water currents. We use adaptive mesh refinement, allowing us to resolve the spatial and temporal scales of wind, current, and wave breaking from $\mathcal{O}(10^{-3})$ m to $\mathcal{O}(10^{0})$ m. Such numerical methodology for two-phase flows, as implemented in Basilisk, has been successfully applied to a range of turbulent two-phase processes, including drops and bubbles in turbulence~\citep{riviere2021sub,perrard2021bubble,farsoiya2023role}, wave breaking~\citep{deike2016air,mostert2022high}, the atmospheric boundary layer~\citep{van2018towards}, and wind-driven wave growth~\citep{wu2021wind,wu2022revisiting}. Here, we extend this framework to wind-forced breaking waves, building on our previous work on momentum fluxes~\citep{scapin2025mom}.
\section{Set-up and summary of the simulated cases}\label{app:cases}
In this section, we recall the two main simulation steps, described in \cite{wu2022revisiting,scapin2025mom} and we provide a summary of the different simulated cases.
\subsection{Initialization and precursor}
The simulations are conducted in two steps: the generation of the precursor air-side turbulent field and the actual two-phase simulation, which resolves the coupled wind, waves, and current. \par
First, we initialize the airflow region using a fully developed turbulent flow field at the desired friction Reynolds number, $Re_{\ast,\lambda}$, following~\citet{wu2022revisiting,scapin2025mom}. This initialization involves a precursor simulation conducted independently in a single-phase setup and employing the same domain. During this precursor simulation, the wave field with profile $\eta_0$ remains at rest, and a no-slip/no-penetration boundary condition is enforced on the wave surface for the velocity field, using the embedded boundary method~\citep{johansen1998cartesian} available within the Basilisk framework~\citep{ghigo2021conservative}. The precursor simulation is performed long enough until the turbulent airflow achieves a statistically steady state by adding an external body force per unit mass acting in the streamwise direction, i.e. $\partial p_0/\partial x(1-\mathcal{H})\mathbf{e}_x$ with $\mathbf{e}_x=(1,0,0)$, on the right-hand side of the momentum equation~\eqref{eqn:momc}. In the expression of the body force, $\partial p_0/\partial x$ is a uniform pressure gradient driving the flow and, here, reads as
\begin{equation}\label{eqn:pif}
  \dfrac{\partial p_0}{\partial x} = \dfrac{\rho_au_\ast^2}{h_a}\mathrm{.}
\end{equation}
The imposed pressure gradient sets the nominal friction velocity $u_\ast$ and prescribes the total stress $\rho_au_\ast^2$ on the wave field. Once a statistically steady state is achieved for the precursor, the resulting fully developed turbulent field is employed as an initial condition for the airflow region in the two-phase simulations. \par 
The second step involves initializing the water orbital velocity using the third-order potential flow solution obtained for a third-order Stokes wave. After the two-phase simulation begins, the airflow and wave field dynamically evolve without any prescribed conditions at the two-phase interface.
\subsection{Summary of the simulated cases}
In table~\ref{tab:set_up}, we report the different simulated cases with the corresponding dimensionless numbers for each case.
\begin{table}[h!]
  \centering
  \begin{tabular}{cccccccc}
    \toprule
    {{$u_\ast/c$}} & {{$Re_{\ast,\lambda}$}} & {{$Re_W$}} & {{$\mu_w/\mu_a$}} & {{$h_a/\lambda$}} & {{$\mathrm{Le}$}} & {{$\mathrm{CPU}$ $\mathrm{hours}$}} \\
    \toprule
    0.30 & 214  & $2.55 \cdot 10^{4}$ & 22.84 & 3.36 & 10 & $4.2\cdot 10^5$ \\
    0.40 & 214  & $2.55 \cdot 10^{4}$ & 17.13 & 3.36 & 10 & $4.2\cdot 10^5$ \\
    0.50 & 214  & $2.55 \cdot 10^{4}$ & 13.71 & 3.36 & 10 & $4.2\cdot 10^5$ \\
    0.70 & 214  & $2.55 \cdot 10^{4}$ & 9.79 & 3.36 & 10 & $4.2\cdot 10^5$ \\
    0.90 & 214  & $2.55 \cdot 10^{4}$ & 7.61 & 3.36 & 10 & $4.2\cdot 10^5$ \\
    \midrule
    0.90 & 53.5 & $2.55 \cdot 10^{4}$ & 1.90 & 3.36 & 10 & $4.2\cdot 10^5$ \\
    0.90 & 107  & $2.55 \cdot 10^{4}$ & 3.81 & 3.36 & 10 & $4.2\cdot 10^5$ \\
    \midrule
    0.90 & 214  & $2.55 \cdot 10^{4}$ & 7.61 & 3.36 & 11 & $1.2\cdot 10^6$ \\
    0.90 & 107  & $2.55 \cdot 10^{4}$ & 3.81 & 6.72 & 11 & $1.2\cdot 10^6$ \\
    \bottomrule
  \end{tabular}
  \caption{Summary of the simulated cases for different values of $u_\ast/c$. In the table, $Re_{\ast,\lambda}=\rho_au_\ast\lambda/\mu_a$, $Re_W=\rho_wc\lambda/\mu_w$, $\mu_w/\mu_a$ and $h_a/\lambda$. The cases with $h_a/\lambda=3.36$ correspond to $4$ waves per box size, whereas the case with $h_a/\lambda=6.72$ to $8$ waves per box size. For the different cases, the initial steepness is set equal to $a_0k=0.3$, and the density ratio is taken as $\rho_w/\rho_a=816$ and the Bond number $Bo=200$.}
  \label{tab:set_up}
\end{table}
Note that the numerical grid in the simulations is adaptive, featuring a minimum grid size $\Delta = L_0/(2^\mathrm{Le})$, where $\mathrm{Le}$ represents the maximum level of refinement. The Adaptive Mesh Refinement (AMR) technique significantly reduces computational costs by maintaining a highly refined grid near the interface and in the boundary layers while allowing coarser grids in the bulk airflow, provided that refinement criteria are met~\citep{popinet2015quadtree,van2018towards}. Following~\citet{wu2022revisiting,scapin2025mom}, the refinement criteria for the air and water velocity components and the volume fraction are set equal to $\epsilon_{ua}=0.3u_\ast$, $\epsilon_{uw} = 10^{-3}c$ and $\epsilon_\mathcal{F}=10^{-4}$, respectively. \par
The computational costs of these simulations include generating a precursor simulation and running the two-phase simulations. Four precursors are used for varying air-side Reynolds number ($Re_{\ast,\lambda}=53.5-107-214$) and the ratio $h_a/\lambda=3.36-6.72$, amounting to $\approx 1.5\cdot 10^5$ CPU each. Each two-phase simulation at $\mathrm{Le}=10$ requires $\approx 4.2\cdot 10^5$ CPU hours, while the two simulations at $\mathrm{Le}=11$ require $\approx 1.20\cdot 10^6$ CPU hours each. We employ $384$ processors for the precursor simulations and for the two-phase cases $480$ at $\mathrm{Le}=10$, and $980$ processors at $\mathrm{Le}=11$. The total cost of the simulation campaign is about $6\cdot 10^6$ CPU hours.
\section{Grid convergence test}\label{app:grid_conv}
\begin{figure*}[h]
  \centering
  \includegraphics[width=\textwidth]{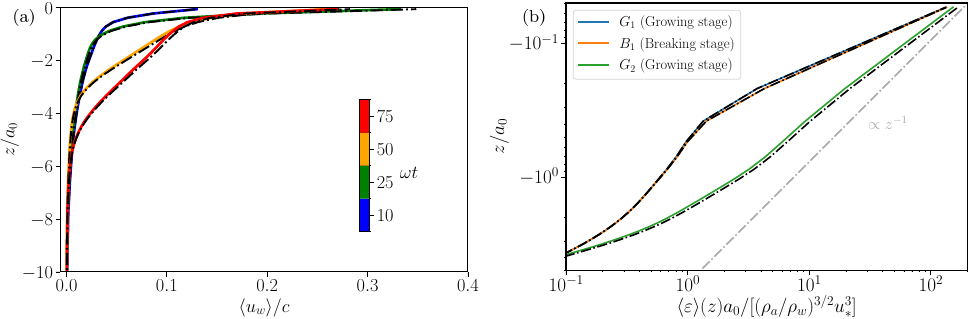}
  \caption{Grid-convergence test for the case $u_\ast/c=0.9$, $Re_{\ast,\lambda}=214$ using two resolution levels: $\mathrm{Le}=10$ (solid colored lines) and $\mathrm{Le}=11$ (dot–dashed lines). The panels display (a) vertical profiles of underwater dissipation at $\omega t = [10,25,50,75]$, and (b) vertical profiles of turbulent dissipation during the first growing ($G_1$), first breaking ($B_1$), and second growing ($G_2$) cycles.}
  \label{fig:grid_conv}
\end{figure*}

In this section, we assess the grid convergence of the simulation for the case $u_\ast/c=0.9$, $Re_{\ast,\lambda}=214$ using two resolution levels, $\mathrm{Le}=10$ and $\mathrm{Le}=11$. Results are shown in figure~\ref{fig:grid_conv}, where we report the vertical profiles of the underwater velocity and the turbulent dissipation at different stages of the wave field. The comparison indicates that $\mathrm{Le}=10$ is sufficient to fully resolve the underwater currents, both in terms of velocity and dissipation. 
\section{A note on the wall-layer scaling for the underwater dissipation}\label{app:alt_scalig}
We briefly summarize why the physical argument behind modeling dissipation using a wall-layer turbulence model is not appropriate for the underwater dissipation due to wave breaking. The wall-layer model assumes a balance between turbulence production by mean shear and its dissipation, as is typical in a turbulent boundary layer. Applying a logarithmic velocity profile leads to the following expression for the turbulence dissipation rate:
\begin{equation}\label{eqn:eps_wl}
  \langle\varepsilon\rangle_{wl}(z) = \dfrac{u_{\ast,w}^3}{\kappa z}\mathrm{,}
\end{equation}
where $u_{\ast,w}$ is the friction velocity in water, and $\kappa = 0.41$ is the von Kármán constant. Following~\citet{terray1996estimates}, $u_{\ast,w}$ is estimated as $u_{\ast,w}=u_\ast\sqrt{\rho_a/\rho_w}$, assuming that the total stress in the air is fully transferred to the water side. \par
\begin{figure*}[h]
  \centering
  \includegraphics[width=\textwidth]{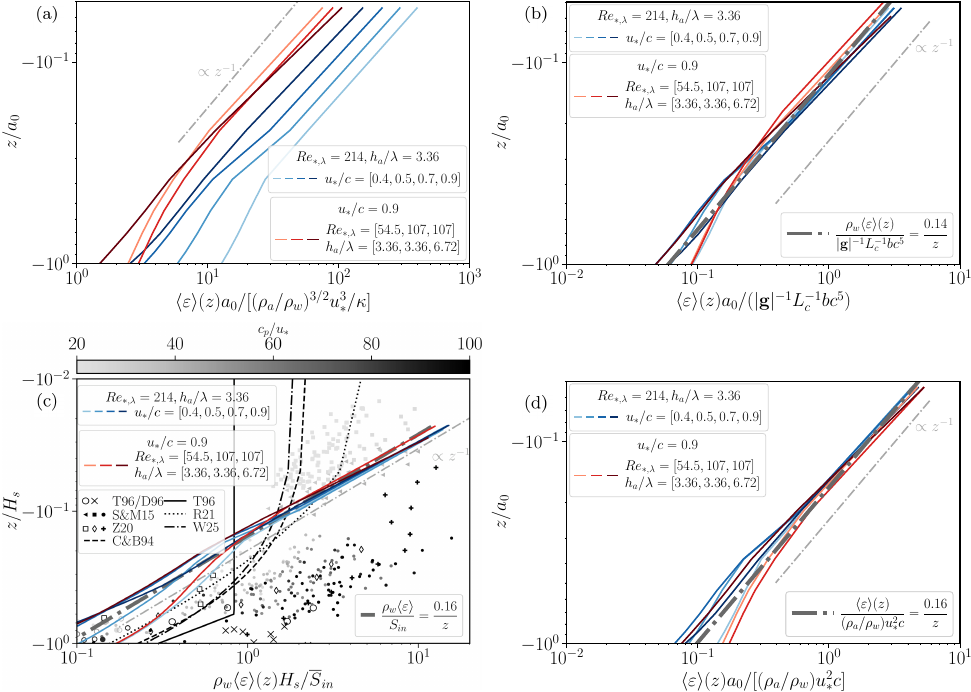}
  \caption{Dissipation profiles after breaking ($G_2$), normalized by: (a) wall-layer scaling $(\rho_a/\rho_w)^{3/2}u_\ast^3/(\kappa a_0)$, (b) the dissipation-based scaling $\overline{S}_{ds}/(\rho_w a_0)$ with $\overline{S}_{ds}=\rho_wg^{-1} L_c^{-1} b c^5$, (c) the wind input $\overline{S}_{in}$, which is extracted from the DNS in the present work, while in~\cite{terray1996estimates,drennan1996oceanic,sutherland2015field}, it is computed using a spectral parametrization, (d) wind-input scaling $(\rho_a/\rho_w)u_\ast^2c/a_0$. In (a)-(b)-(d), the depth is normalized by the initial amplitude $a_0$, in (c) by the significant wave height $H_s$ to compare with field data~\citep{terray1996estimates,drennan1996oceanic,sutherland2015field}.}
  \label{fig:wl_Sin}
\end{figure*}

We test the wall-layer scaling using the dissipation profiles from the second growth cycle but find that it fails to collapse the cases for different values of $u_\ast/c$, as clearly shown by figure~\ref{fig:wl_Sin}(a). This discrepancy arises because the assumed production-dissipation balance is disrupted by wave breaking, which substantially enhances turbulence dissipation compared to the estimate of equation~\eqref{eqn:eps_wl}. This enhancement has been demonstrated in several previous studies~\citep{craig1994modeling,gemmrich1994energy,terray1996estimates,sutherland2015field,zippel2020measurements} and is also demonstrated in figure 3 of the main text. \par
A more physically representative scaling is derived in the main text, based on the energy dissipation due to breaking $\overline{S}_{ds}=\rho_wg^{-1}L_c^{-1}bc^5$, which we repeat in this supplementary in figure~\ref{fig:wl_Sin}(b) or based on wind input given the balance between dissipation due to breaking and wind input (shown in main text and repeated in figure\ref{fig:wl_Sin}(c)). We note that a simplified scaling based on wind input can be proposed since the wind input can be approximated as $\overline{S}_{in} = \rho_a u_\ast^2 c$ and the vertical profile can be rescaled as
\begin{equation}\label{eqn:eps_s}
  \langle\varepsilon\rangle(z) = \tilde{A} \dfrac{\rho_a}{\rho_w} \dfrac{u_\ast^2 c}{z}\mathrm{,}
\end{equation}
where $\tilde{A} \approx 0.16$ is a non-dimensional constant obtained by best fit to the data. This value is very similar to the constant $A$ used in figure~\ref{fig:wl_Sin}(c) for the group $\rho_w \langle\varepsilon\rangle(z)H_s/ \overline{S}_{in}$ and to the constant $\mathcal{A}$ used in figure~\ref{fig:wl_Sin}(b) for the group $\langle\varepsilon\rangle(z) / (|\mathbf{g}|^{-1} L_c^{-1} b c^5)$. Note that equation~\eqref{eqn:eps_s} incorporates the effects of both the friction velocity $u_\ast$ and the wave phase speed $c$.
Figure~\ref{fig:wl_Sin}(d) shows the dissipation profiles normalized using the scaling in equation~\eqref{eqn:eps_s}, revealing excellent collapse within a water depth of $5a_0$ for a range of $u_\ast/c$ and $Re_{\ast,\lambda}$ values. \par
It is important to emphasize that the scaling in equation~\eqref{eqn:eps_s} is fully consistent with the dissipation-based scaling in equation 14 of the main text, given the balance between wind input $\overline{S}_{in}$ and energy dissipation $\overline{S}_{ds}$.
%
%
\bibliographystyle{apacite}
\bibliography{agubib_sup}